\documentclass[preprint,5p,twocolumn,12pt]{elsarticle}

\usepackage{graphicx}

\usepackage{amssymb}
 
 \usepackage{amsmath}

\biboptions{sort&compress}

\usepackage{color}

\journal{Journal of Colloid and Interface Science}

\begin{document}

\begin{frontmatter}



\title{Ion size effects on the electrokinetics \\ of salt-free concentrated suspensions in ac fields}


\author[FA1]{Rafael Roa}
\author[FA1]{F\'elix Carrique\corref{carrique}}
\author[FA2]{Emilio Ruiz-Reina}

\address[FA1]{F\'isica Aplicada I, Universidad de M\'alaga, 29071, M\'alaga (Spain)}
\address[FA2]{F\'isica Aplicada II, Universidad de M\'alaga, 29071, M\'alaga (Spain)}

\cortext[carrique]{carrique@uma.es}

\begin{abstract}
We analyze the influence of finite ion size effects in the response of a salt-free concentrated suspension of spherical particles to an oscillating electric field. Salt-free suspensions are just composed of charged colloidal particles and the added counterions released by the particles to the solution, that counterbalance their surface charge. In the frequency domain, we study the dynamic electrophoretic mobility of the particles and the dielectric response of the suspension. We find that the Maxwell-Wagner-O'Konski process associated with the counterions condensation layer, is enhanced for moderate to high particle charges, yielding an increment of the mobility for such frequencies. We also find that the increment of the mobility grows with ion size and particle charge. All these facts show the importance of including ion size effects in any extension attempting to improve standard electrokinetic models.

\end{abstract}


\end{frontmatter}




\section{Introduction} \label{sec:intro}

In the last years, there has been a renewed interest in electrophoresis. This is in part due to recent advances in nanoscience which make possible the separation of macromolecules by size or charge. Suspended DNA or proteins are driven and separated by applying dc or ac electric fields \cite{Lavrentovich2010,Sparreboom2009,Dorfman2010}. Another main application in the field of nanoscience is the use of gold nanoparticles for drug delivery or cancer cell detection \cite{Giljohann2010,Dukhin2010}. Measurements of the electrophoretic mobility have been found to be useful to characterize the surface functionalization of these gold nanoparticles \cite{LopezViota2009}. Usually particles are charged and suspended together with microions and a structure of electric double layer (EDL) appears around the particle \cite{Dukhin1974,Lyklema1995}. The electrophoretic mobility of a suspended particle is not only dependent on the particle charge or the viscosity of the medium, but also on the configuration of the EDL. 

Most of the theoretical EDL models are based on the classical Poisson-Boltzmann equation, a mean field approach that takes into account point-like ions in solution. This theory breaks down when the crowding of ions becomes significant, and steric repulsion and correlations potentially become important. Some authors have shown that the consideration of finite ion size effects allows for the crowding of ions near the particle surface \cite{Bazant2009}. This redistribution of ions modifies the EDL around the particle and consequently its electrophoretic mobility when an external electric field is applied \cite{ArandaRascon2009,Khair2009,LopezGarcia2011,Roa2011c}.

We can find in the literature different studies dealing with ion size effects. Some of them concern microscopic descriptions of ion-ion correlations \cite{Lobaskin2007,Pagonabarraga2010}. These approaches are mainly restricted to equilibrium conditions, but are able to predict important phenomena like overcharging \cite{Lyklema2009}. Other studies are based on macroscopic descriptions considering average interactions by mean field approximations \cite{ArandaRascon2009,Bikerman1942,Borukhov1997,Roa2011,Bazant2011}. In many of these works, the finite ion size is commonly included by modifying the activity coefficient of the ions in the electrochemical potential or by incorporating entropic contributions related to the excluded volume of the ions. The macroscopic approaches have been found to work appreciably well with monovalent electrolytes for high particle charges and/or large ionic sizes when they have been compared with some simulation results \cite{IbarraArmenta2009}.

Most works in electrokinetics concern suspensions with low particle concentration, but nowadays it is the concentrated regime that deserves more attention because of its practical applications. These systems are difficult to understand due to the inherent complexity associated with the increasing particle-particle electrohydrodynamic interactions as particle concentration grows \cite{OBrien1990,Dukhin2010b}. On the other hand, systems with low salt concentration show a lower screening of the repulsive electrostatic particle-particle interactions, which favors the generation of colloidal crystals or glasses. Suspensions just composed of charged particles and their ionic countercharges (the so-called added counterions) in the liquid medium are named salt-free suspensions. The interest in these systems has increased in recent years from both experimental and theoretical points of view \cite{Medebach2003,Palberg2004,Reinmuller2010,Ohshima2003,Chiang2006,Carrique2008,Carrique2008b,Bastea2010}. 

The dielectric response of colloidal suspensions as a function of the frequency of the applied external electric field is a powerful tool. Its analysis provides rich information on the dynamics of the EDL because it is very sensitive to the particle-solution interface. Carrique \textit{et al.} have studied the dynamic properties of salt-free concentrated suspensions with point-like ions when an ac electric field is applied \cite{Carrique2008,Carrique2008b}. To our knowledge, the only theoretical work in the frequency domain considering ion size effects is the one of Aranda-Rasc\'on \textit{et al.} for dilute suspensions with electrolytes \cite{ArandaRascon2009b}. The same authors have shown that the consideration of a minimum approach distance of ions to the particle surface, not necessarily equal to their effective radius in the bulk solution, can predict overcharging for high electrolyte concentrations and counterion valence \cite{LopezGarcia2011}.

Our aim in this paper is to analyze the influence of finite ion size effects in the response of a salt-free concentrated suspension to an oscillating electric field. We will study specially the dynamic electrophoretic mobility of the particles and the dielectric response of the suspension in the frequency domain. Recently, we have studied the equilibrium EDL \cite{Roa2011} and the response to a static electric field \cite{Roa2011c} of this kind of suspensions with ion size effects. In this paper, we will extend our previous works to ac electric fields following the treatment of Carrique \textit{et al.} for salt-free concentrated suspensions with point-like ions \cite{Carrique2008,Carrique2008b}.  

The plan of the paper is as follows. In Section \ref{sec:model} we describe the electrokinetic model to account for ion size effects. We give details of the resolution method of the electrokinetic equations and define the quantities we calculate in Section \ref{sec:method}. The results of the numerical calculations are shown in Section \ref{sec:results} and analyzed upon changing particle surface charge density, particle volume fraction, and size of the counterions. In order to show the realm of the finite ion size effect in salt-free suspensions, the results are compared with standard predictions for point-like ions. Conclusions are presented in Section \ref{sec:conclusions}.


\section{Model} \label{sec:model}

\subsection{Electrokinetic equations}

We use a cell model \cite{Kuwabara1959,Zholkovskij2007} to study the macroscopic properties of the suspension from appropriate averages of local properties in a representative cell. In this approach, each spherical particle of radius $a$ is surrounded by a concentric shell of the liquid medium, having an outer radius $b$ such that the particle/cell volume ratio in the cell is equal to the particle volume fraction throughout the entire suspension, that is
\begin{equation}\label{phidef}
\phi=\left( \frac{a}{b}\right)^3
\end{equation}

In this approximation, we simulate the hydrodynamic and electrical interactions between particles in the suspension by proper specification of boundary conditions at the outer surface of the cell.

Let us consider a spherical charged particle of radius $a$, surface charge density $\sigma$, mass density $\rho_p$ and relative permittivity $\epsilon_{rp}$ immersed in a salt-free medium of relative permittivity $\epsilon_{rs}$, mass density $\rho_s$ and viscosity $\eta$, with only the presence of the added counterions of valence $z_c$ and drag coefficient $\lambda_c$. We consider finite size counterions as spheres of radius $R$ with a point charge at their center. By applying to the system an oscillating electric field $\mathbf{E}\ \mathrm{e}^{-i\omega t}$ of angular frequency $\omega$, the particle moves with a velocity $\mathbf{v}_e\mathrm{e}^{-i\omega t}$, the dynamic electrophoretic velocity. The axes of the spherical coordinate system ($r$, $\theta$, $\varphi$) are fixed at the center of the particle, with the polar axis ($\theta=0$) parallel to the electric field. The solution of the problem at time $t$ requires the knowledge, at every point $\mathbf{r}$ of the system, of the electric potential, $\Psi(\mathbf{r},t)$, the number density of counterions, $n_c(\mathbf{r},t)$, their drift velocity, $\mathbf{v}_c(\mathbf{r},t)$, the fluid velocity, $\mathbf{v}(\mathbf{r},t)$, and the pressure, $P(\mathbf{r},t)$. The electrokinetic equations connecting them are \cite{Dukhin1974,Mangelsdorf1992}: \begin{equation}\label{poisson}
  \nabla^2\Psi(\mathbf{r},t)=-\frac{z_ce}{\epsilon_0\epsilon_{rs}}n_c(\mathbf{r},t)
\end{equation}
\begin{multline}
  \eta\nabla^2\mathbf{v}(\mathbf{r},t)-\nabla P(\mathbf{r},t)-z_cen_c(\mathbf{r},t)\nabla \Psi(\mathbf{r},t)\\ 
  =\rho_s \frac{\partial}{\partial t}[\mathbf{v}(\mathbf{r},t)+\mathbf{v}_e\ \mathrm{e}^{-i\omega t}]
\end{multline}
\begin{equation}
  \nabla\cdot[n_c(\mathbf{r},t) \mathbf{v}_c(\mathbf{r},t)]=-\frac{\partial}{\partial t}[n_c(\mathbf{r},t)]
\end{equation}
\begin{multline}\label{nersnt-p}
  n_c(\mathbf{r},t)\mathbf{v}_c(\mathbf{r},t)=n_c(\mathbf{r},t)\mathbf{v}(\mathbf{r},t)\\ -\frac{1}{\lambda_c}n_c(\mathbf{r},t)\nabla\mu_c(\mathbf{r},t)
\end{multline}
\begin{equation}\label{incompress}
  \nabla\cdot\mathbf{v}(\mathbf{r},t)=0
\end{equation}

In these equations, $\mu_c(\mathbf{r},t)$ is the electrochemical potential of the counterions, $\epsilon_0$ is the vacuum permittivity and $e$ is the elementary electric charge. The drag coefficient $\lambda_c$ is related to the diffusion coefficient by $\lambda_c=k_BT/D_c$, where $k_B$ is Boltzmann's constant, and $T$ is the absolute temperature.

As we are interested in studying the linear response of the system to an electric field, we apply a perturbation scheme. Thus, each quantity $X$ is written as the sum of its equilibrium value, $X^0$, plus a perturbation term, $\delta X$, linearly dependent with the field multiplied by the term $\mathrm{e}^{-i\omega t}$, that represents the time dependent sinusoidal response of the stationary state \cite{Carrique2008}.

We introduce the finite size of the counterions by considering their excluded volume and including the entropy of the solvent molecules in the free energy of the suspension, $F=U-TS$ \cite{Roa2011}
\begin{multline}
  U=\int \mathrm{d}\textbf{r}\bigg{[}-\frac{\epsilon_0\epsilon_{rs}}{2}|\nabla\Psi^0(\mathbf{r})|^2\\ +z_cen_c^0(\mathbf{r})\Psi^0(\mathbf{r})-\mu_c^0n_c^0(\mathbf{r})\bigg{]}
\end{multline}
\begin{multline}\label{entropy}
  -TS=k_BTn_c^{max}\int \mathrm{d}\textbf{r}\bigg{[}\frac{n_c^0(\mathbf{r})}{n_c^{max}}\ln \left(\frac{n_c^0(\mathbf{r})}{n_c^{max}}\right)\\ 
  +\left(1-\frac{n_c^0(\mathbf{r})}{n_c^{max}}\right)\ln \left(1-\frac{n_c^0(\mathbf{r})}{n_c^{max}}\right)\bigg{]}
\end{multline}
being $n_c^{max}$ the maximum possible concentration of counterions due to the excluded volume effect, defined as $n_c^{max}=V^{-1}$, where $V$ is the average volume occupied by an ion in the solution. The last term in Eq. (\ref{entropy}) is the one that accounts for the ion size effect, and was proposed earlier by Borukhov et al. \cite{Borukhov1997}. Performing the variations of the free energy with respect to $\Psi^0(\mathbf{r})$ and $n_c^0(\mathbf{r})$, combining both resulting expressions and applying spherical symmetry, we obtain
\begin{multline}\label{mpb0}
\frac{\mathrm{d}^2\Psi^0(r)}{\mathrm{d}r^2}+\frac{2}{r}\frac{\mathrm{d}\Psi^0(r)}{\mathrm{d}r}\\ =-\frac{z_ce}{\epsilon_0\epsilon_{rs}}\frac{b_c\exp\left(-\frac{z_ce\Psi^0(r)}{k_BT}\right)}{1+\frac{b_c}{n_c^{max}}\left[\exp\left(-\frac{z_ce\Psi^0(r)}{k_BT}\right)-1\right]}
\end{multline}
where $b_c$ is an unknown coefficient that represents the ionic concentration where the equilibrium electric potential is chosen to be zero. Details about this modified Poisson-Boltzmann equation including ion size effects can be found in Ref. \cite{Roa2011}.

To obtain the perturbation terms of the quantities of interest, due to the symmetry of the problem, we make use of the following spherical functions: $h(r)$, $\phi_c(r)$, and $Y(r)$ \cite{Ohshima1997}
\begin{multline}\label{vOhshima}
  \mathbf{v}(\mathbf{r})=(v_r,v_\theta,v_\varphi)=\\ \left(-\frac{2}{r}h(r)E\cos\theta,\frac{1}{r}\frac{\mathrm{d}}{\mathrm{d}r}(rh(r))E\sin\theta,0\right)
\end{multline}
\begin{equation}\label{deltamuc}
  \delta\mu_c(\mathbf{r})=-z_ce\phi_c(r)E\cos\theta
\end{equation}
\begin{equation}\label{deltapsi}
  \delta\Psi(\mathbf{r})=-Y(r)E\cos\theta
\end{equation}
with $E=|\mathbf{E}|$.

Substituting the above mentioned perturbation scheme into the differential electrokinetic equations, Eqs. (\ref{poisson})-(\ref{incompress}), neglecting nonlinear perturbations terms, and making use of the symmetry conditions of the problem we obtain
\begin{multline}\label{L_navier-stokes}
  \mathcal{L}(\mathcal{L}h(r))+\frac{i\omega\rho_s}{\eta}\mathcal{L}h(r)=-\frac{z_ce^2}{k_BT\eta r}\\ \times\left(\frac{\mathrm{d}\Psi^0(r)}{\mathrm{d}r}\right)n_c^0(r)\left(\phi_c(r)-\frac{n_c^0(r)}{n_c^{max}}Y(r)\right)
\end{multline}
\begin{multline}\label{L_continuity}
  \mathcal{L}\phi_c(r)+\frac{i\omega\lambda_c}{k_BT}\left(\phi_c(r)-Y(r)\right)=\frac{e}{k_BT}\left(\frac{\mathrm{d}\Psi^0(r)}{\mathrm{d}r}\right)\\ \times \left(1-\frac{n_c^0(r)}{n_c^{max}}\right)\left(z_c\frac{\mathrm{d}\phi_c(r)}{\mathrm{d}r}-\frac{2\lambda_c}{e}\frac{h(r)}{r}\right)
\end{multline}
\begin{equation}\label{L_poisson}
  \mathcal{L}Y(r)=-\frac{z_c^2e^2n_c^0(r)}{\epsilon_0\epsilon_{rs}k_BT}\left(\phi_c(r)-Y(r)\right)
\end{equation}
where the $\mathcal{L}$ operator is defined by
\begin{equation}
  \mathcal{L} \equiv \frac{\mathrm{d}^2}{\mathrm{d}r^2}+\frac{2}{r}\frac{\mathrm{d}}{\mathrm{d}r}-\frac{2}{r^2}
\end{equation}

In the case of a static electric field, $\omega=0$, Eqs. (\ref{L_navier-stokes})-(\ref{L_poisson}) turn into the expressions obtained in Ref. \cite{Roa2011c}. For point-like counterions, $n_c^{max}=\infty$, these equations become those of Refs. \cite{Carrique2008,Carrique2008b}.

According to Ref. \cite{ArandaRascon2009}, we incorporate a distance of closest approach of the counterions to the particle surface, resulting from their finite size. We assume that the counterions cannot come closer to the surface of the particle than their effective hydration radius, $R$, and, therefore, the ionic concentration will be zero in the region between the particle surface, $r=a$, and the spherical surface, $r=a+R$, defined by the counterion effective radius. This reasoning implies that counterions are considered as spheres of radius $R$ with a point charge at their center. 

With this consideration, we solve the electrokinetic equations, Eqs. (\ref{mpb0}), (\ref{L_navier-stokes})-(\ref{L_poisson}), only between $r=a+R$ and $r=b$. When we address the problem in the region between $r=a$ and $r=a+R$, the equations to solve turn into the Laplace equation for the equilibrium electric potential, and equations $\mathcal{L}(\mathcal{L}h(r))=0$, $\phi_c(r)=0$, and $\mathcal{L}Y(r)=0$ for the rest of the spherical functions, because this region is free of charge. We call FIS+L model this complete model that includes ion size effects and also considers the distance of closest approach of the counterions to the charged particle surface.

\subsection{Boundary conditions}

The boundary conditions needed to solve the electrokinetic equations are analogous, but dealing with complex quantities, to those described in Ref. \cite{Roa2011c} (Section II B) for the response of a salt-free concentrated suspension to a static electric field including ion size effects. At the particle surface, we apply the continuity of the electric potential, the discontinuity of the normal component of the displacement vector, the non-slip condition for the fluid and the impenetrability of ions to the solid surface. On the outer surface of the cell, we use the Kuwabara's boundary conditions for the fluid velocity field and the Shilov-Zharkikh-Borkovskaya conditions for the perturbed electric potential. Finally, if we consider a distance of closest approach of the counterions to the particle surface, we also need the continuity of the pressure and both normal and tangential components of the fluid velocity and the vorticity. A remarkable difference with the static case is that the net force acting on the particle or the unit cell is not zero. Details of the net force calculation can be found in Ref. \cite{Ohshima1997} or Appendix 1 in Ref. \cite{Carrique2008}.

In terms of the radial functions $\Psi^0(r)$, $Y(r)$, $\phi_c(r)$ and $h(r)$, the boundary conditions are:

\noindent
(i) at the particle surface $r=a$
\begin{equation}
\frac{\mathrm{d}\Psi^0(r)}{\mathrm{d}r}\bigg |_{r=a}=-\frac{\sigma}{\epsilon_0\epsilon_{rs}}
\end{equation}
\begin{equation}
  \frac{\mathrm{d}Y(r)}{\mathrm{d}r}\bigg |_{r=a}-\frac{\epsilon_{rp}}{\epsilon_{rs}}\frac{Y(a)}{a}=0
\end{equation}
\begin{equation}
h(a)=0
\end{equation}
\begin{equation}
\frac{\mathrm{d}h(r)}{\mathrm{d}r}\bigg |_{r=a}=0
\end{equation}
(ii) at the surface $r=a+R$ defined by the counterion effective radius
\begin{equation}
\Psi^0(a+R^-)=\Psi^0(a+R^+)
\end{equation}
\begin{equation}
\frac{\mathrm{d}\Psi^0(r)}{\mathrm{d}r}\bigg |_{r=a+R^-}=\frac{\mathrm{d}\Psi^0(r)}{\mathrm{d}r}\bigg |_{r=a+R^+}
\end{equation}
\begin{equation}
Y(a+R^-)=Y(a+R^+)
\end{equation}
\begin{equation}
\frac{\mathrm{d}Y(r)}{\mathrm{d}r}\bigg |_{r=a+R^-}=\frac{\mathrm{d}Y(r)}{\mathrm{d}r}\bigg |_{r=a+R^+}
\end{equation}
\begin{equation}
\frac{\mathrm{d}\phi_c(r)}{\mathrm{d}r}\bigg |_{r=a+R^+}=0
\end{equation}
\begin{equation}
h(a+R^-)=h(a+R^+)
\end{equation}
\begin{equation}
\frac{\mathrm{d}h(r)}{\mathrm{d}r}\bigg |_{r=a+R^-}=\frac{\mathrm{d}h(r)}{\mathrm{d}r}\bigg |_{r=a+R^+}
\end{equation}
\begin{equation}
\mathcal{L}h(a+R^-)=\mathcal{L}h(a+R^+)
\end{equation}
\begin{multline}
  \frac{\mathrm{d^3}h(r)}{\mathrm{d}r^3}\bigg |_{r=a+R^-}=\frac{\mathrm{d^3}h(r)}{\mathrm{d}r^3}\bigg |_{r=a+R^+}\\  -\frac{z_ce}{(a+R)\eta}n_c^0(a+R^+)Y(a+R^+)
\end{multline}
(iii) and finally, at the outer surface of the cell $r=b$
\begin{equation}
\Psi^0(b)=0
\end{equation}
\begin{equation}
\frac{\mathrm{d}\Psi^0(r)}{\mathrm{d}r}\bigg |_{r=b}=0
\end{equation}
\begin{equation}
Y(b)=b
\end{equation}
\begin{equation}
\phi_c(b)=b
\end{equation}
\begin{equation}
\mathcal{L}h(b)=0
\end{equation}
\begin{multline}\label{netforce}
\eta\frac{\mathrm{d}}{\mathrm{d}r}\big[r\mathcal{L}h(r)\big]_{r=b}-z_ceb_cY(b)=i\omega\rho_s
\\\times\left(h(b)-2\phi\frac{\rho_p-\rho_s}{\rho_s}h(b)-b\frac{\mathrm{d}h(r)}{\mathrm{d}r}\bigg |_{r=b}\right)
\end{multline}

This last boundary condition, Eq. (\ref{netforce}), stands for the equation of motion of the unit cell. In the case of a static electric field, $\omega=0$, we recover the expression for the net force showed in Ref. \cite{Roa2011c}.


\section{Method and calculated quantities} \label{sec:method}
\subsection{Method}

We will discuss the results of the proposed FIS+L electrokinetic model. In order to show the realm of the finite ion size effect in salt-free suspensions, the results are compared with standard predictions for point-like ions, PL model \cite{Carrique2008,Carrique2008b}. The electrokinetic equations with their boundary conditions form a boundary value problem that can be solved numerically using the MATLAB routine bvp4c \cite{Kierzenka2001}.

\begin{table}[b]
\centering
\caption{Parameter values used in the calculations.}
\label{table-parameters}
\renewcommand{\tabcolsep}{0.4cm}
\begin{tabular}{ll}
\\ \hline 
$T=298.15$ K & $a=100$ nm  \\
$\eta=0$.89$\cdot10^{-3}$ P  & $z_c=+1$ \\
$\epsilon_{rs}=$ 78.55 & $D_c=9$.34$\cdot10^{-9}$ m$^2$/s \\
$\epsilon_{rp}=2$ & \\
\hline
\end{tabular}
\end{table}

For the sake of simplicity, we assume that the average volume occupied by a counterion is $V=(2R)^{3}$, being $2R$ the counterion effective diameter. With this consideration, the maximum possible concentration of counterions due to the excluded volume effect is $n_c^{max}=(2R)^{-3}$. This corresponds to a simple cubic package (52\% packing). In molar concentrations, the values used in the calculations, $n_c^{max}=$ 22, 4 and 1.7 M, correspond approximately to counterion effective diameters of $2R=$ 0.425, 0.75 and 1 nm, respectively. These are typical hydrated ionic diameters \cite{Israelachvili1992}. We present in Table \ref{table-parameters} the parameter values used in all the calculations. The chosen parameters correspond to hydrated H$^+$ counterions, which are commonly found in many experimental conditions with salt-free suspensions of, for example, negatively charged sulfonated polymer particles, due to the cleaning process of the suspension with proton exchange resins.

\subsection{Calculated quantities}

The dynamic electrophoretic mobility $\mu$ of a spherical particle in a concentrated colloidal suspension can be defined from the relation between the electrophoretic velocity of the particle and the macroscopic electric field. According to Refs. \cite{Roa2011c,Carrique2008} it can be determined through
\begin{equation}
\mu=\frac{2h(b)}{b}
\end{equation}

We calculate the nondimensional dynamic electrophoretic mobility as
\begin{equation}
\mu^*=\frac{3\eta e}{2\epsilon_0\epsilon_{rs}k_BT}\mu
\end{equation}

The complex conductivity, $K$, of the suspension is usually defined in terms of the volume averages of the local electric current density and electric field in a cell representing the whole suspension. Following a similar procedure to that described for the dc conductivity in Ref. \cite{Roa2011c}, we obtain (see also Ref. \cite{Carrique2008b})
\begin{multline}
K=\left(\frac{z_c^2e^2}{\lambda_c}\frac{\mathrm{d}\phi_c(r)}{\mathrm{d}r}\bigg |_{r=b}-\frac{2h(b)}{b}z_ce\right)n_c^0(b)\\-i\omega\epsilon_{rs}\epsilon_0\frac{\mathrm{d}Y(r)}{\mathrm{d}r}\bigg |_{r=b}
\end{multline}

From the complex conductivity, the real $\epsilon_r'(\omega)$ and imaginary $\epsilon_r''(\omega)$ components of the complex relative permittivity of the suspension $\epsilon_r(\omega)$ are calculated by writing
\begin{multline}
K(\omega)=K(\omega=0)-i\omega\epsilon_0\epsilon_r(\omega)\\
=K(\omega=0)+\omega\epsilon_0\epsilon_r''(\omega)-i\omega\epsilon_0\epsilon_r'(\omega)
\end{multline}
\begin{equation}
\epsilon_r'(\omega)=-\frac{\mathrm{Im}[K(\omega)]}{\omega\epsilon_0}
\end{equation}
\begin{equation}
\epsilon_r''(\omega)=\frac{\mathrm{Re}[K(\omega)]-K(\omega=0)}{\omega\epsilon_0}
\end{equation}


\section{Results and discussion} \label{sec:results}

\subsection{Point-like model}\label{subsec:PLmodel}

The classical frequency response of a salt-free concentrated suspension with point-like counterions, PL model, is as follows: (i) at low frequency, the electromigration and diffusion processes have enough time to be fully developed around the particle and, commonly, this fact leads to the generation of an induced electric dipole moment that tends to brake the particle motion. In this frequency region, there is a plateau value of the dynamic electrophoretic mobility that coincides with the electrophoretic mobility in static electric fields. (ii) As the frequency increases, we find a frequency region where the counterions cannot follow the comparatively fast field oscillations. Thus, the above mentioned dipolar moment decreases and, consequently, the dynamic mobility increases. This process is known as Maxwell-Wagner-O'Konski (MWO) relaxation and takes place whenever the medium and the charged particle, surrounded by its EDL, present different conductivities and permittivities. (iii) Finally, the frequency can be so high that the inertia of the particle and fluid restricts the motion progressively. As a result, the mobility shows a continuous decline when the frequency rises, which is known as the inertial relaxation.

There is another classical relaxation mechanism, the alpha relaxation  \cite{Dukhin1974}, which is related to the concentration polarization effect (\textit{i.e.}, the presence of a gradient of neutral electrolyte around the particle). We do not find any alpha relaxation in a salt-free suspension, as was explained in Ref. \cite{Carrique2010}, because we only have one ionic species, the added counterions.  

In order to clarify the discussion, we will try to separate the different mechanisms by performing three different variations of the PL model:

(i) the complete PL model that includes all the mentioned effects;

(ii) the  \textit{pure inertial response} where we do not allow any perturbation of the ionic atmosphere from the equilibrium values. So, breaking mechanisms associated with the charge polarization are excluded in this variation;

(iii) the \textit{inertia-free response}, where we have eliminated all the inertial terms in the electrokinetic equations.

\begin{figure}[t]
\centering
  \includegraphics[width=8.5cm]{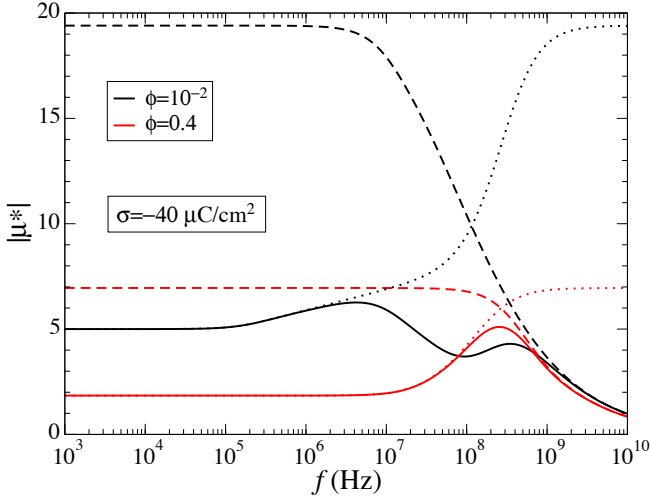}
  \caption{Modulus of the scaled dynamic electrophoretic mobility as a function of frequency for different particle volume fractions. We use three variations of the PL model. Solid lines stand for results of the complete PL model. Dashed lines show the \textit{pure inertial response}. Dotted lines display the \textit{inertia-free response}.}
  \label{fgr:inertia}
\end{figure}

In Fig. \ref{fgr:inertia} we show the modulus of the scaled dynamic mobility for the three last-mentioned PL variations. We use two different particle volume fractions at a given particle charge density. We display in solid lines the complete PL model, in dashed lines the  \textit{pure inertial response}, and in dotted lines the \textit{inertia-free response}.

The  \textit{pure inertial response} behaves as follows: after an initial low frequency mobility plateau, the mobility monotonously decreases with frequency. This plateau has larger values than that of the complete PL model. The difference between the \textit{pure inertial response} and the PL model is due to the absence of breaking effects on particle motion associated with the induced dipole moment (double layer relaxation effect).

In the numerical results corresponding to the \textit{inertia-free response}, we observe one or two successive increments in the dynamic mobility for high or low particle volume fraction, respectively. These increments are related to one or two successive MWO relaxation processes in each case. When we introduce the inertia (dashed lines) to get the complete PL model (solid lines), these above mentioned increments become into one or two successive peaks, as can be observed. The diminution of the dynamic mobility after the first maximum is therefore due to the inertial relaxation.

As was expected, once all the MWO processes have relaxed in the high frequency limit, the mobility in the \textit{inertia-free response} curve reaches the same plateau value than the one of the \textit{pure inertial response} for low frequencies. This is because, when the induced polarization completely disappears, the counterions distribution coincides with that of the equilibrium as in the \textit{pure inertial response}.

\subsection{Maxwell-Wagner-O'Konski relaxations}\label{subsec:MWO}

We have observed that two differenced MWO relaxations can exist. They will be related to two differenced regions in the EDL. Through the Wagner formula for a constant dielectric mixture, it is possible to obtain the frequency and the dielectric increment of a MWO relaxation \cite{Dukhin1974}
\begin{equation}\label{freq-MWO}
\omega_\mathrm{MWO}=\frac{(1-\phi)K_p+(2+\phi)K_s}{(1-\phi)\epsilon_0\epsilon_{rp}+(2+\phi)\epsilon_0\epsilon_{rs}}
\end{equation}
\begin{multline}\label{de-MWO}
\Delta\epsilon_\mathrm{MWO}=\frac{9\phi(1-\phi)}{(1-\phi)\epsilon_{rp}+(2+\phi)\epsilon_{rs}} \\ \times \left[\frac{\epsilon_{rs}K_p-\epsilon_{rp}K_s}{(1-\phi)K_p+(2+\phi)K_s}\right]^2
\end{multline}
where $K_p$ and $K_s$ are the conductivities of the particles and the medium, respectively. The particles are assumed to be made of a nonconducting material. Their conductivity is exclusively associated with the surface conductivity, $K^\sigma$, that appears due to an excess of counterions in the ionic atmosphere, $K_p=2K^\sigma/a$ \cite{Lyklema1995}. Eqs. (\ref{freq-MWO}) and (\ref{de-MWO}) were derived without allowance of mutual polarization of particles and they are valid for suspensions with added electrolyte, thin EDL, and reasonably low $\phi$.

The latter equations predict only one MWO relaxation process. However, for suspensions of highly charged particles, two different MWO relaxations have been considered in the literature  to explain their dielectric response. This consideration is based on the existence of two differenced regions in the EDL, specially when $\sigma$ is sufficiently high and ion size effects are considered, see Fig. 2 on Ref. \cite{Roa2011}. The first one is a condensate of counterions very close to the particle surface. The second one is a diffuse layer that extends from the end of the condensate to the outer surface of the cell. When we have finite size counterions, the condensate consists in a homogeneous region where counterions are well packaged. In the PL case, such picture of the condensate is not valid, but there are theoretical evidences of the existence of a thin region with different behavior in the electric potential and ionic distribution than in the diffuse layer \cite{Ohshima2002,Carrique2010}. As was suggested on Ref. \cite{Carrique2010}, we will consider each region with a different associated MWO relaxation process and roughly estimate their MWO relaxation frequencies and dielectric increments to qualitatively explain the behavior of both MWO relaxation processes.

To obtain the relaxation frequency of the condensate, we need to calculate the surface conductivity of the counterion condensation layer. Considering this layer with a mean concentration $n_c^{max}$ and a thickness $\delta$, we obtain
\begin{equation}
K^\sigma=\frac{z_c^2e^2n_c^{max}\delta}{\lambda_c}
\end{equation}

For the study of the condensate relaxation, the conductivity of the counterions in the diffuse layer, $K_s$, will be taken equal to zero, because it has no influence in the condensate relaxation process. Using $K_p=2K^\sigma/a$ and $\epsilon_{rp}\ll\epsilon_{rs}$, according to Eq. (\ref{freq-MWO}) we obtain a relaxation frequency
\begin{equation}\label{freq-cond}
\omega_\mathrm{MWO}^{cond}=\frac{2z_c^2e^2n_c^{max}\delta(1-\phi)}{\epsilon_0\epsilon_{rs}a\lambda_c(2+\phi)}
\end{equation}
and with Eq. (\ref{de-MWO}) a dielectric increment
\begin{equation}\label{de-cond}
\Delta\epsilon_\mathrm{MWO}^{cond}=\frac{9\phi\epsilon_{rs}}{(1-\phi)(2+\phi)}
\end{equation}

To obtain only the relaxation process of the diffuse layer, we need to calculate the conductivity of the counterions in this region, $K_s$,
\begin{equation}
K_s=\frac{z_c^2e^2\overline{n}_c}{\lambda_c}=\frac{-3z_ce\sigma^{dif}\phi}{a\lambda_c(1-\phi)}
\end{equation}
where $\overline{n}_c$ is the average counterions concentration in the diffuse layer, and $\sigma^{dif}$ is the charge density at the spherical surface, $r=a+\delta$, defined by the thickness of the condensation layer. We now take $K_p=0$ because we consider the particle with its condensate as an equivalent particle with less surface charge density, $\sigma^{dif}$. Introducing the expression of the conductivity $K_s$ in Eq. (\ref{freq-MWO}) and considering $\epsilon_{rp}\ll\epsilon_{rs}$, we obtain the relaxation frequency of the diffuse layer
\begin{equation}\label{freq-dif}
\omega_\mathrm{MWO}^{dif}=\frac{-3z_ce\sigma^{dif}\phi}{\epsilon_0\epsilon_{rs}a\lambda_c(1-\phi)}
\end{equation}
and with Eq. (\ref{de-MWO}) the dielectric increment
\begin{equation}\label{de-dif}
\Delta\epsilon_\mathrm{MWO}^{dif}=\frac{9\phi(1-\phi)\epsilon_{rp}^2}{(2+\phi)^3\epsilon_{rs}}
\end{equation}

As commented before, the previous expressions will be more precise for reasonably low $\phi$. Also, the expression for the MWO relaxation frequency of the condensate will be more accurate than the one for the diffuse layer: the use of an average concentration works better in the condensate because it is a thin layer with homogeneous ionic density.

\subsection{Finite ion size}\label{subsec:fis}

We will jointly study both the dynamic electrophoretic mobility of the particles and the dielectric response of the suspension as a function of frequency, because they are strongly interrelated.

\subsubsection{Condensate MWO relaxation}

\begin{figure}[t]
\centering
  \includegraphics[width=8.45cm]{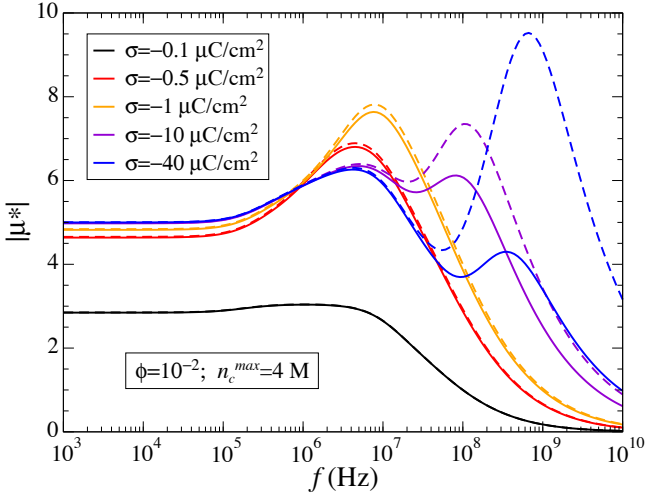}
  \caption{Modulus of the scaled dynamic electrophoretic mobility of the particles as a function of frequency for different particle surface charge densities. All calculations performed at low particle volume fraction. Solid lines show the results for point-like ions. Dashed lines show the results of the FIS+L model with $n_c^{max}=4$ M.}
  \label{fgr:mobpdsigma-dil}
\end{figure}

\begin{figure}[t]
\centering
  \includegraphics[width=8.45cm]{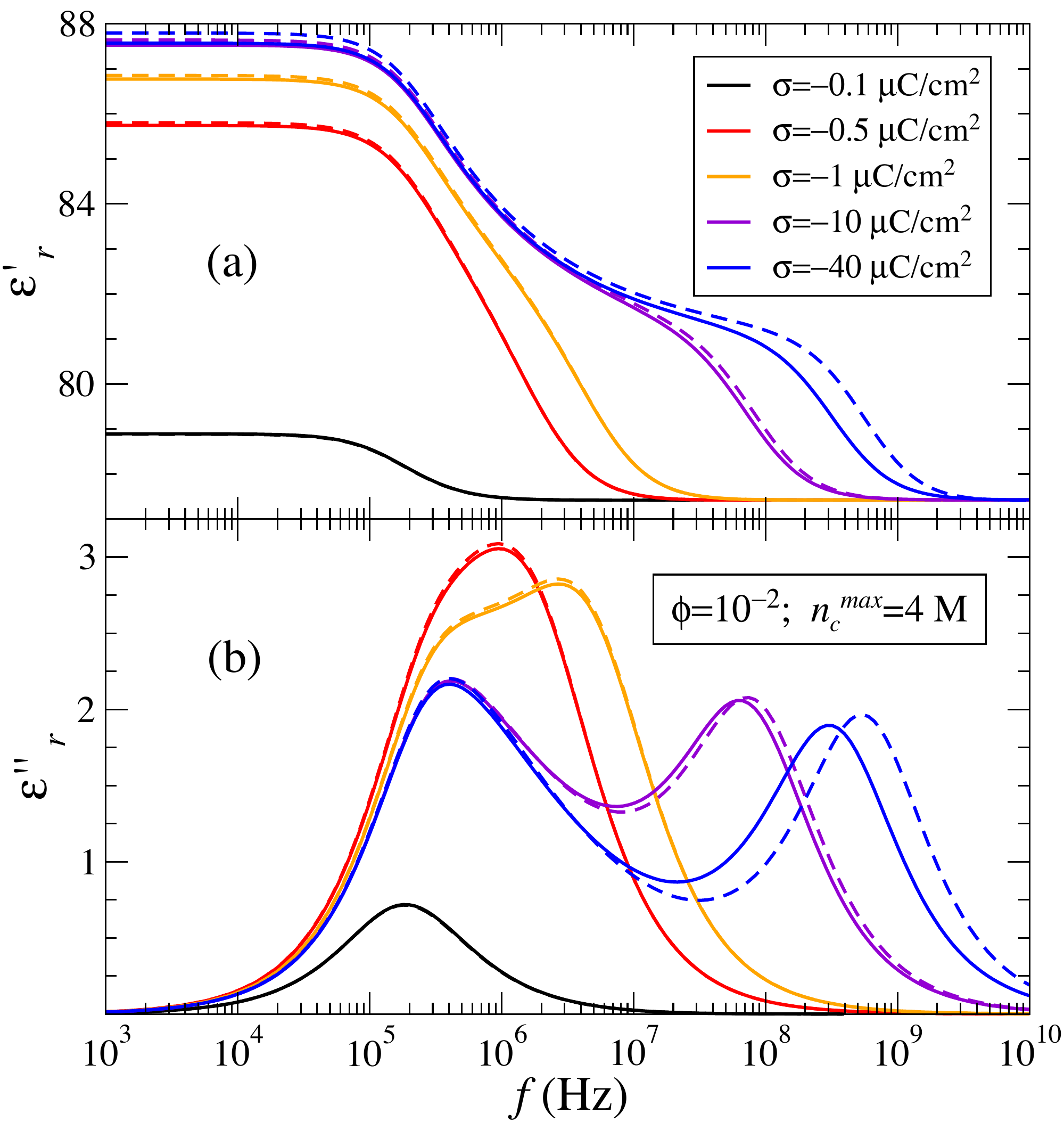}
  \caption{Real (a) and imaginary (b) parts of the relative permittivity of the suspension as a function of frequency for different particle surface charge densities. All calculations performed at low particle volume fraction. Solid lines show the results for point-like ions. Dashed lines show the results of the FIS+L model with $n_c^{max}=4$ M.}
  \label{fgr:diepdsigma-dil}
\end{figure}

Fig. \ref{fgr:mobpdsigma-dil} represents the modulus of the scaled dynamic electrophoretic mobility and Fig. \ref{fgr:diepdsigma-dil} the real (a) and the imaginary (b) parts of the relative permittivity of a salt-free concentrated suspension as a function of frequency. We compare the results of the FIS+L model (dashed lines) at a fixed counterion size, $n_c^{max}=4$ M, with those of the PL model (solid lines). Different colors stand for different particle surface charges. The calculations were made at low particle volume fraction, $\phi=10^{-2}$.

For the low frequency mobility and permittivity plateaus, Figs. \ref{fgr:mobpdsigma-dil} and \ref{fgr:diepdsigma-dil}a, we observe that there is almost no difference between the finite size and point-like results for any particle charge at low particle volume fraction. As was discussed in Ref. \cite{Roa2011c} for the static mobility and conductivity, this is because in the case of a dilute suspension the inclusion of ion size effects only significantly modifies the counterions fluxes in the immediate vicinity of the particle.

The MWO relaxation frequency is defined as that of the maximum in the imaginary part of the permittivity, as it is well-known. We note that for a suspension with low volume fraction, depending on particle surface charge, one or two differenced MWO relaxations, Fig. \ref{fgr:diepdsigma-dil}, or analogously one or two mobility maximums, Fig. \ref{fgr:mobpdsigma-dil}, may take place. As we indicated before, we associate the first one upon increasing frequency with the MWO relaxation of the diffuse part of the EDL, and the second one with the relaxation of the condensate. 

At low $\sigma$ there is no condensate of counterions near the particle surface and therefore no condensate MWO relaxation process is observed. When we rise the particle charge, almost all the extra counterions accumulate in the condensate \cite{Ohshima2002}, which seriously grows when also ion size effects are considered \cite{Roa2011}. This explains why ion size effects do not produce any remarkable effect in the MWO relaxation of the diffuse part of the EDL and why they considerably enhance the condensate MWO relaxation. 

According to Eq. (\ref{freq-dif}), the frequency of the MWO of the diffuse layer changes with $\sigma^{dif}$. As the counterions concentration in the diffuse layer has been scarcely altered, $\omega_\mathrm{MWO}^{dif}$ remains basically the same in Figs. \ref{fgr:mobpdsigma-dil} and \ref{fgr:diepdsigma-dil}. On the contrary, Fig. \ref{fgr:diepdsigma-dil}b shows an increment of the frequency of the condensate MWO relaxation when ion size effects are considered. This is in agreement with Eq. (\ref{freq-cond}), because $\omega_\mathrm{MWO}^{cond}$ increases when the width of the condensate, $\delta$, raises. For the well resolved MWO peaks of the two highest surface charge curves in Fig. \ref{fgr:diepdsigma-dil}b, we observe that the height of the peaks is nearly independent on both, the surface charge density and the ion size, in accordance with Eqs. (\ref{de-cond}) and (\ref{de-dif}).

In a previous paper, we studied the effects of the electric polarization on the magnitude of the static electrophoretic mobility in a salt-free concentrated suspension with finite ion size effects \cite{Roa2011c}. This study was based on a procedure developed by Bradshaw-Hajek \textit{et al.} \cite{Bradshaw2010}. We showed that the induced charge polarization density was larger when ion size effects were considered. The generalization of the latter study to ac electric fields leads to similar conclusions. As the relaxation effect and, correspondingly, the induced dipole moment, have been increased with ion size, the second mobility maximum in Fig. \ref{fgr:mobpdsigma-dil} attains higher values as well. This is due to the disappearance of breaking mechanisms on the particle motion of increasing importance as the ion size grows, causing the dynamic mobility to reach superior values when the size of the counterions is taken into account.

\subsubsection{Overlapping of MWO relaxations}

\begin{figure}[t]
\centering
  \includegraphics[width=8.5cm]{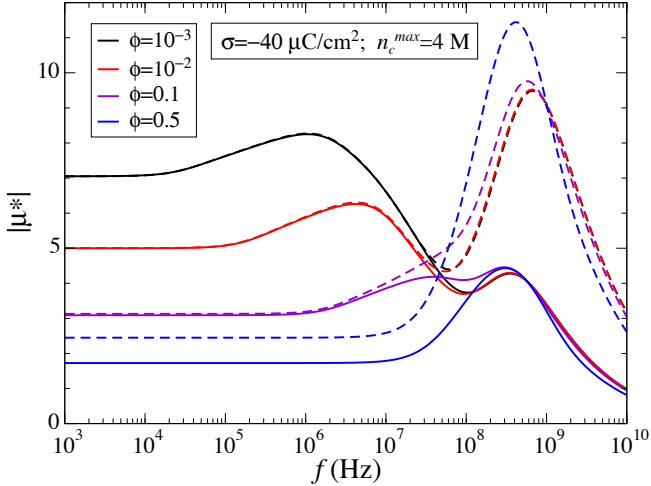}
  \caption{Modulus of the scaled dynamic electrophoretic mobility of the particles as a function of frequency for different particle volume fractions. Solid lines show the results for point-like ions. Dashed lines show the results of the FIS+L model with $n_c^{max}=4$ M.}
  \label{fgr:mobsdphi}
\end{figure}

\begin{figure}[t]
\centering
  \includegraphics[width=8.5cm]{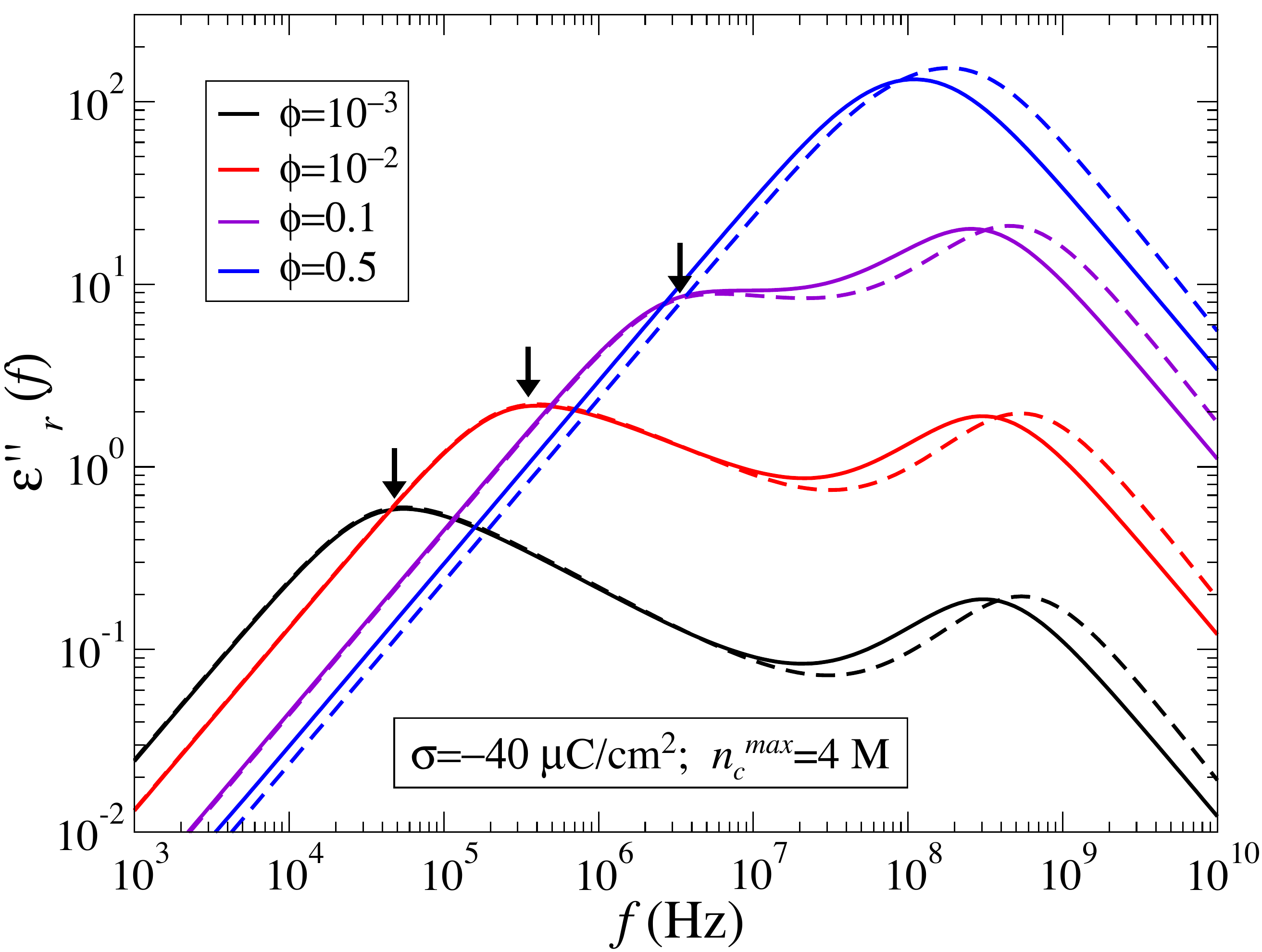}
  \caption{Imaginary part of the relative permittivity of the suspension as a function of frequency for different particle volume fractions. Solid lines show the results for point-like ions. Dashed lines show the results of the FIS+L model with $n_c^{max}=4$ M.}
  \label{fgr:diesdphi}
\end{figure}

We show the frequency response of the modulus of the scaled dynamic electrophoretic mobility and the imaginary part of the relative permittivity of the suspension in Figs. \ref{fgr:mobsdphi} and \ref{fgr:diesdphi}, respectively. In both Figures we compare the results of the FIS+L model with $n_c^{max}=4$ M, dashed lines, with those of the PL model, solid lines, for different particle volume fractions. All the calculations were performed at a high particle charge, $\sigma=-40\ \mu$C/cm$^2$.

We see in Figs. \ref{fgr:mobpdsigma-dil} and \ref{fgr:diepdsigma-dil} how two differenced MWO relaxations take place when surface charge increases in conditions of low volume fraction. Now we observe that the MWO relaxations of the condensate and the diffuse layer in Fig. \ref{fgr:diesdphi}, or analogously the two mobility maximums in Fig. \ref{fgr:mobsdphi}, tend to overlap in frequency for concentrated suspensions at high surface charge. According to Eq. (\ref{freq-dif}), $\omega_\mathrm{MWO}^{dif}$ grows with volume fraction at a rate $\phi/(1-\phi)$. This is in agreement with the frequency shift observed in the MWO relaxation of the diffuse layer, indicated with black arrows in Fig. \ref{fgr:diesdphi}. Eq. (\ref{freq-cond}) predicts a frequency change with volume fraction at a rate $(1-\phi)/(2+\phi)$ for the condensate MWO relaxation. Then we find no significant changes in $\omega_\mathrm{MWO}^{cond}$ for low $\phi$ values, and a small decrease for high volume fractions as shown in Fig. \ref{fgr:diesdphi}. These behaviors result in the observed overlapping of the MWO relaxations for concentrated suspensions.

When we include ion size effects we find only changes in the condensate MWO relaxation (enhancement of the corresponding mobility maximum and small increase in $\omega_\mathrm{MWO}^{cond}$). These changes can be explained with the same reasoning used for Figs. \ref{fgr:mobpdsigma-dil} and \ref{fgr:diepdsigma-dil}: the consideration of finite size counterions seriously enlarges the condensate near the particle but does not produce remarkable effects in the diffuse layer. We also observe the well-known diminution of mobility with the increase of volume fraction in Fig. \ref{fgr:mobsdphi} due basically to the larger screening of the particle charge \cite{Roa2011}: when the particle concentration grows, the available space for the counterions inside the cell decreases and, consequently, the screening of the particle charge is greatly raised, thus reducing the value of the surface potential and, therefore, the mobility.

\subsubsection{Highly charged concentrated suspensions}

\begin{figure}[t]
\centering
  \includegraphics[width=8.5cm]{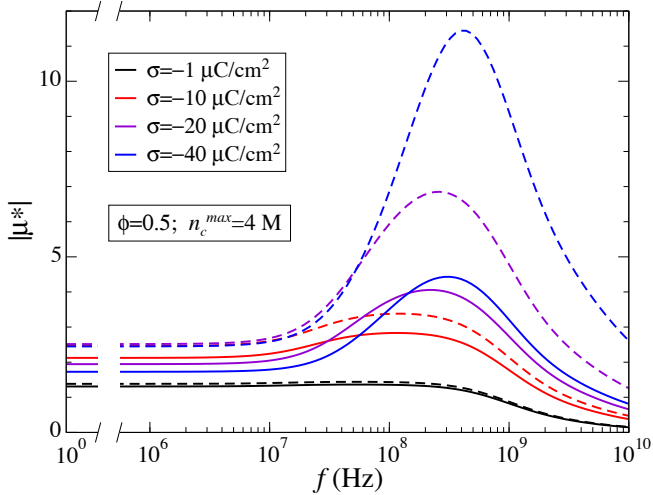}
  \caption{Modulus of the scaled dynamic electrophoretic mobility of the particles as a function of frequency for different particle surface charge densities. All calculations performed at high particle volume fraction. Solid lines show the results for point-like ions. Dashed lines show the results of the FIS+L model with $n_c^{max}=4$ M.}
  \label{fgr:mobpdsigma}
\end{figure}

\begin{figure}[t]
\centering
  \includegraphics[width=8.5cm]{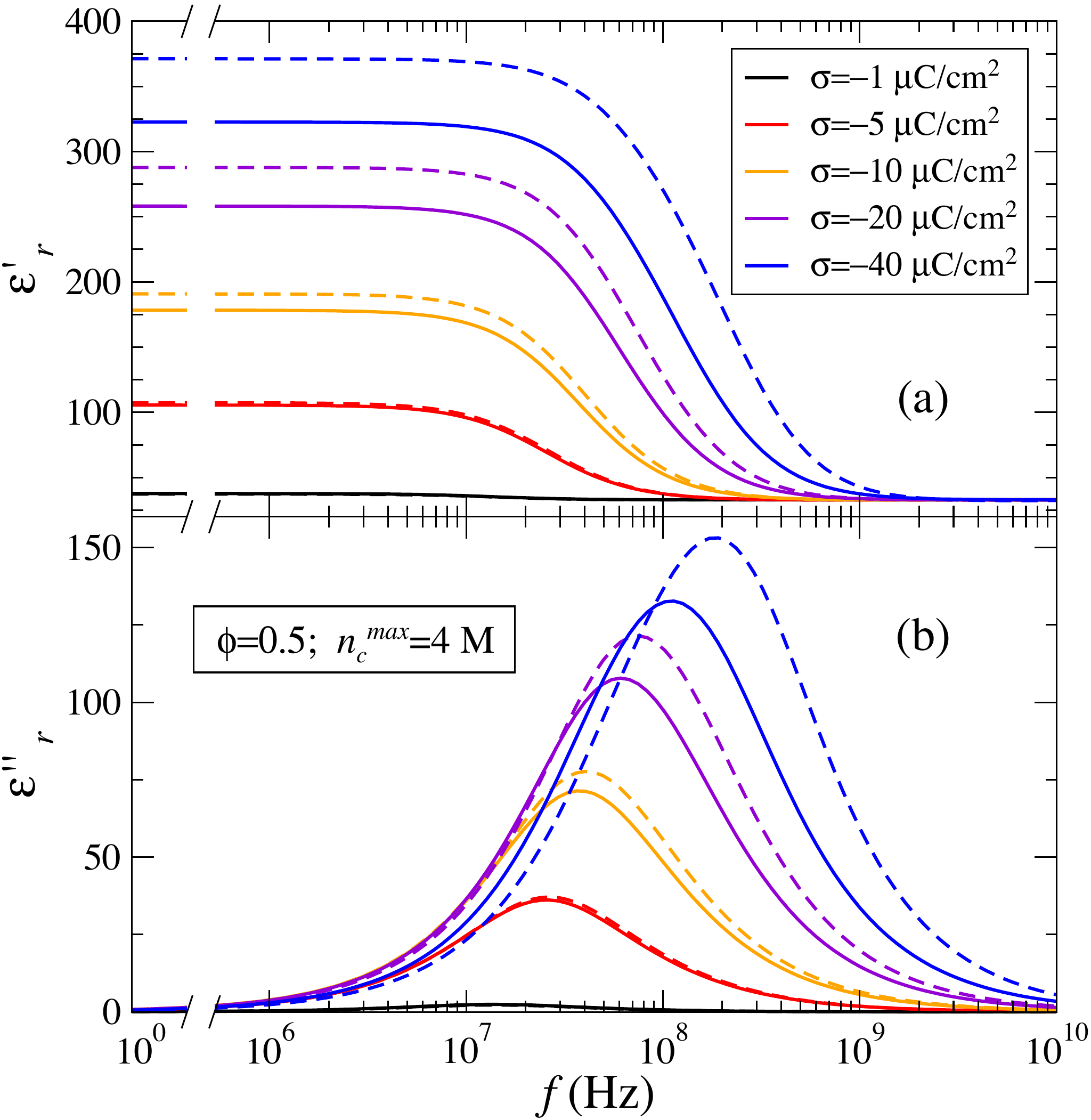}
  \caption{Real (a) and imaginary (b) parts of the relative permittivity of the suspension as a function of frequency for different particle surface charge densities. All calculations performed at high particle volume fraction. Solid lines show the results for point-like ions. Dashed lines show the results of the FIS+L model with $n_c^{max}=4$ M.}
  \label{fgr:diepdsigma}
\end{figure}

\begin{figure}[t]
\centering
  \includegraphics[width=8.5cm]{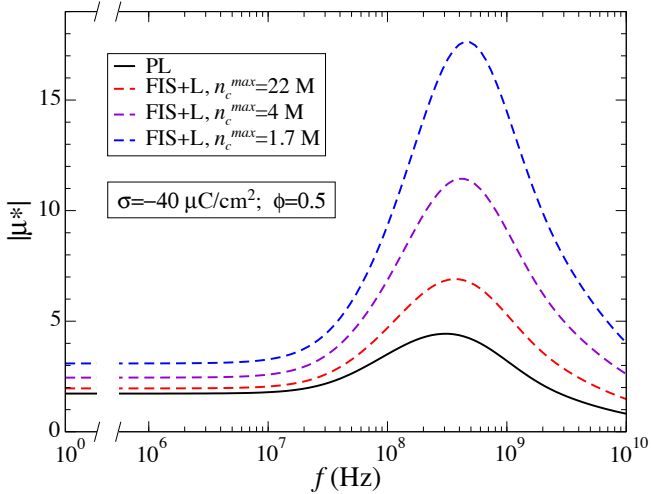}
  \caption{Modulus of the scaled dynamic electrophoretic mobility of the particles as a function of frequency for different ion sizes (dashed lines). Black lines show the results for point like ions.}
  \label{fgr:mobs40p05}
\end{figure}

\begin{figure}[t]
\centering
  \includegraphics[width=8.5cm]{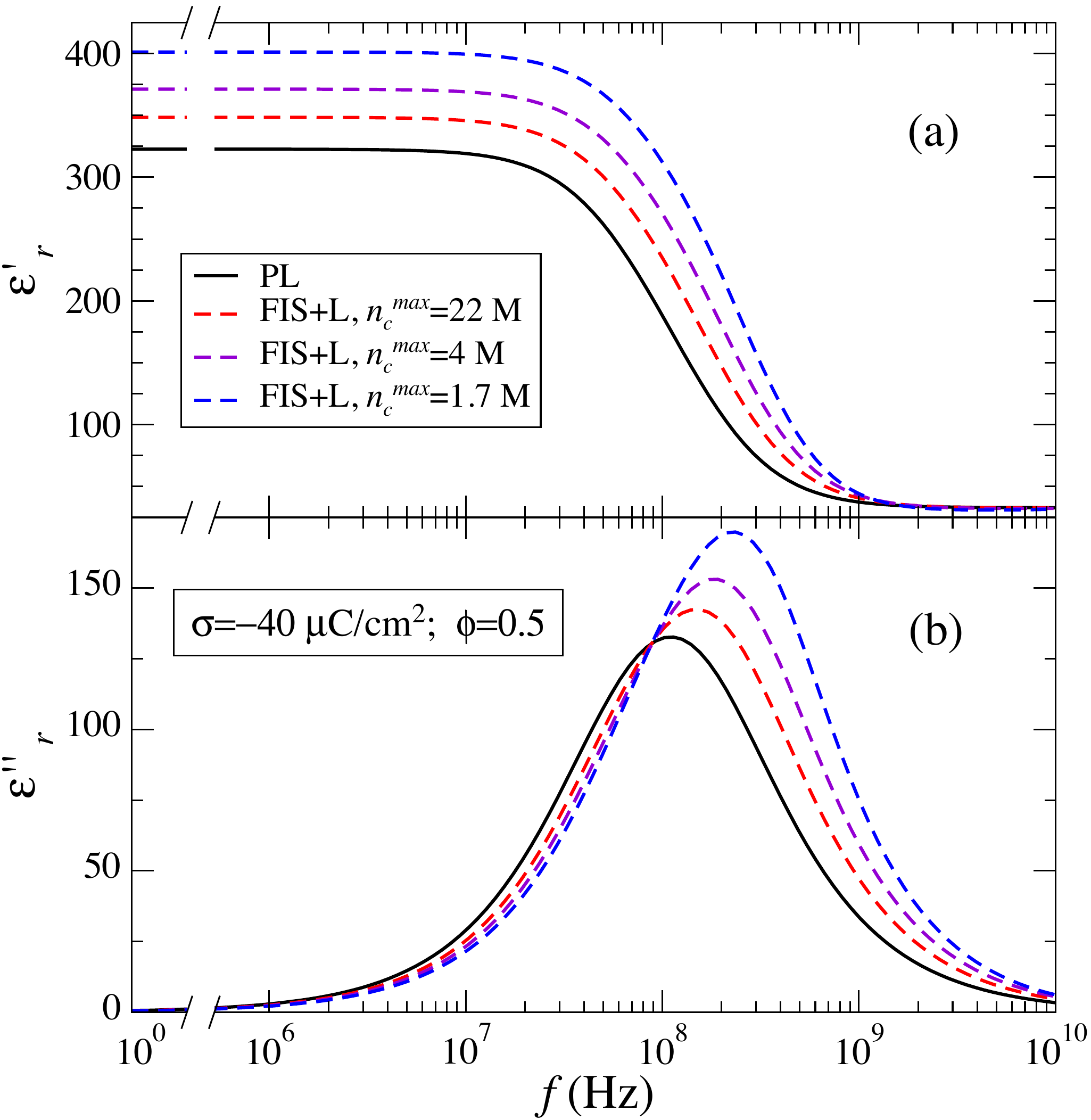}
  \caption{Real (a) and imaginary (b) parts of the relative permittivity of the suspension as a function of frequency for different ion sizes (dashed lines). Black lines show the results for point like ions.}
  \label{fgr:dies40p05}
\end{figure}

Figs. \ref{fgr:mobpdsigma} and \ref{fgr:diepdsigma} show the modulus of the scaled dynamic electrophoretic mobility and the real (a) and imaginary (b) parts of the relative permittivity of the suspension as a function of frequency for different particle surface charges. We compare the results of the FIS+L model with $n_c^{max}=4$ M, dashed lines, with those of the PL model, solid lines at a high particle volume fraction, $\phi=0.5$.

Besides, Figs. \ref{fgr:mobs40p05} and \ref{fgr:dies40p05} represent the same quantities at fixed particle surface charge, $\sigma=-40\ \mu$C/cm$^2$, and volume fraction, $\phi=0.5$. In these Figures, we compare the results of the FIS+L model at different ion sizes (different colored dashed lines) with those of the PL model (black lines).

As we mentioned before, for this high volume fraction value, the two MWO relaxations are overlapped in a unique broad peak. We observe how both the dynamic mobility and relative permittivity increase when we consider finite size counterions in comparison with the PL case and when we increase the particle surface charge. The reason is that the increase of the surface charge or the consideration of ion size effects leads to an enhancement of the overall charge polarization in the EDL, resulting in both, higher permittivity values as can be seen in Figs. \ref{fgr:diepdsigma}a and \ref{fgr:dies40p05}a, and larger heights of the corresponding peaks of the imaginary part, Figs. \ref{fgr:diepdsigma}b and \ref{fgr:dies40p05}b. A similar explanation applies to the remarkable increment observed in the mobility maxima, associated with the MWO relaxations, Figs. \ref{fgr:mobpdsigma} and \ref{fgr:mobs40p05}:  the disappearance of the augmented induced dipole moment gives rise to greater mobility values.

Figs. \ref{fgr:diepdsigma}b and \ref{fgr:dies40p05}b display a shift to larger frequencies in the MWO relaxation when ion size effects are considered. This shift is also larger the larger the size of the counterions (lower $n_c^{max}$ value). The MWO relaxation of the diffuse layer is nearly independent of the ion size, Eq. (\ref{freq-dif}), and, therefore, the shift observed is entirely due to the displacement of the condensate MWO relaxation. We have checked numerically that the expression of the MWO relaxation frequency of the condensate, Eq. (\ref{freq-cond}), predicts a shift to higher frequencies with the increase of the ion size through the product $n_c^{max}\delta$. This is because the width of the condensate $\delta$ augments in a higher rate than the parameter $n_c^{max}$ diminishes upon increasing ion size \cite{Roa2011}. When we increase the particle surface charge, we are also increasing the surface conductivity $K^\sigma$ of the condensate, for both PL and FIS+L models, and consequently the frequency of the MWO relaxation rises.


\section{Conclusions} \label{sec:conclusions}

By using a cell model approach we have analyzed the influence of finite ion size effects in the response of a salt-free concentrated suspension of spherical particles to an oscillating electric field. We have derived a mean-field ac electrokinetic model that accounts for the excluded volume of the counterions. 

In the frequency domain, we have studied the dynamic electrophoretic mobility of the particles and the dielectric response of the suspension. For this purpose we have performed a comparative study of the different physical mechanisms, \textit{pure inertia response} and charge polarization relaxations, to know how they interplay to give the complete response. This study has allowed us to characterize the relative importance and relaxation frequencies of each mechanism separately. In the discussion of the numerical results two different MWO relaxations have been successfully associated with the relaxations of the different ionic processes that take place in the diffuse and condensate regions of the EDL. Furthermore, the inclusion of ion size effects leads to an overall increment of the dynamic mobility and relative permittivity in comparison with the point-like case. The enhancement of the MWO relaxation for moderate to high particle charges, which is associated with the counterions condensation layer, has yielded a remarkable increment of the mobility for such frequencies. In addition, we have found that this increment of the mobility grows with ion size and particle charge. Besides, we have observed a shift in the MWO relaxation of the condensate to larger frequencies with ion size.

Some of these calculations can be compared with experimental results. To perform such comparisons, concentrated suspensions of highly charged particles are required. These suspensions have been classically  difficult to synthesize, although existing highly charged sulfonated polystyrene latexes are good candidates. Moreover, high-frequency experimental setups are needed to work with the very high frequency region where the MWO relaxation of the condensate takes place.

%


\section*{Acknowledgements}
Financial support by Junta de Andaluc\'ia, Spain (Project P08-FQM-3779), and MICINN, Spain (Project FIS2010-18972), co-financed with ERDF funds by the EU, is acknowledged.






\bibliographystyle{model1-num-names}
\bibliography{<your-bib-database>}



\end{document}